\documentclass[preprint2]{aastex}

\slugcomment{In Press, Astrophys. and Space Sci., 2012}

\shorttitle{Spectral Content of $^{22}$Na/$^{44}$Ti}
\shortauthors{O'Keefe et al.}

\begin{document}


\title{Spectral Content of $^{22}$Na/$^{44}$Ti Decay Data: Implications for a Solar Influence}


\author{D. O'Keefe, B. L. Morreale and R. H. Lee}
\affil{Physics Department; U.S. Air Force Academy; 2354 Fairchild Dr., USAFA, CO 80840 USA}

\author{John B. Buncher}
\affil{Physics Department Wittenberg University, Springfield, OH 45501 USA}

\author{Ephraim Fischbach, T. Gruenwald, and J. H. Jenkins\altaffilmark{1,2}}
\affil{Department of Physics, Purdue University, West Lafayette, IN 47907 USA}

\author{D. Javorsek II}
\affil{412th Test Wing, Edwards AFB, CA 93524 USA}

\and

\author{P. A. Sturrock}
\affil{Center for Space Science and Astrophysics, Stanford University, Stanford, CA 94305 USA}


\altaffiltext{1}{School of Nuclear Engineering, Purdue University, 400 Central Dr., West Lafayette, IN  47907 USA}
\altaffiltext{2}{Corresponding author: jere@purdue.edu, +1 765 496 3573 (voice), +1 765 494 9570 (fax)}


\begin{abstract}
We report a reanalysis of data on the measured decay rate ratio $^{22}$Na/$^{44}$Ti which were originally published by Norman et al., and interpreted as supporting the conventional hypothesis that nuclear decay rates are constant and not affected by outside influences. We find upon a more detailed analysis of both the amplitude and the phase of the Norman data that they actually favor the presence of an annual variation in $^{22}$Na/$^{44}$Ti, albeit weakly. Moreover, this conclusion holds for a broad range of parameters describing the amplitude and phase of an annual sinusoidal variation in these data. The results from this and related analyses underscore the growing importance of phase considerations in understanding the possible influence of the Sun on nuclear decays. Our conclusions with respect to the phase of the Norman data are consistent with independent analyses of solar neutrino data obtained at Super-Kamiokande-I and the Sudbury Neutrino Observatory (SNO).
\end{abstract}





\section{Introduction}
\label{intro}

Unexplained periodic variations in measured nuclear decay rates have been reported recently by a number of groups in experiments with a variety of detector types and isotopes. These reports, along with the observation of a change in the decay rate of $^{54}$Mn during a solar flare \citet{jen09a} suggest the possibility of a direct solar influence on nuclear decay rates through an as yet unknown mechanism. Periodicities and other ``non-random'' behaviors have been reported in the decays of $^{3}$H \citep{fal01,lob99,vep12}, $^{32}$Si, $^{36}$Cl \citep{alb86,jen09b,jav10,stu10a,stu11a,stu11b,jen12}, $^{54}$Mn \citep{jen11}, $^{56}$Mn \citep{ell90}, $^{60}$Co \citep{bau07,par10a,par10b}, $^{90}$Sr \citep{par10a,par10b,stu12a}, $^{137}$Cs \citep{bau07}, $^{152}$Eu \citep{sie98}, $^{222}$Rn (and/or its daughters) \citep{ste11,stu12b}, and $^{226}$Ra (and/or its daughters) \citep{sie98,jen09b,jav10,stu10b,stu11a,stu11b,fis09}. Since these fluctuations have been seen by groups located at various sites employing different detector technologies (e.g., gas, scintillation, solid state), it is unlikely that they can all be attributed to temperature, pressure, humidity or other ``environmental'' influences on the detector systems. This conclusion is bolstered by two additional observations: A detailed analysis \citep{jen10} of possible environmental influences on the detectors used in the Brookhaven National Laboratory (BNL) \citep{alb86} and the Physikalisch-Technische Bundesanstalt (PTB) \citep{sie98} experiments concluded that such influences were at least an order of magnitude too small to account for the observed periodic effects. Moreover, the changes seen over the short duration of a solar flare \citep{jen09a} could not have been generated by ``seasonal'' influences. Significantly, additional periodicities of approximately 32 d and 173 d associated with internal solar rotation and oscillations have been reported, along with the annual periodicities, in a number of experiments and analyses \citep{stu11a,stu11b,par10b,stu10b,shn98a,shn98b,bau07,stu10a}. Since there are no known environmental effects that can produce such periodic variations in the efficiencies of existing detectors, the implication is that these periodicities reflect an influence external to the Earth environment, such as a solar influence on the decay process itself as suggested in Refs. \citep{jen09a,jen09b,stu11a,stu11b,stu10a,stu10b}.

Other groups have questioned the conclusion that there could be an influence on terrestrial decay rates by a particle or field emanating from the Sun. These groups have reported results from analyses of decay data from experiments based both on Earth and in space, and do not find such an influence. Cooper \citep{coo09} analyzed data from the radioisotope thermoelectric generators (RTGs) onboard the Cassini spacecraft enroute to Saturn and found no evidence for a dependence of the decay rate of $^{238}$Pu (which powers the RTGs) on the Cassini-Sun separation. Our group has recently carried out a more detailed analysis of the RTG data, and we have shown that in fact there are no conflicts between the Cassini results and the previously reported decay anomalies \citep{kra12}. Additionally, it was noted that even if the RTG data analyzed were sufficiently good to support the claims in Ref. \citep{coo09}, there would be no conflict in principle between the absence of an apparent solar influence in the $\alpha$-decay of $^{238}$Pu, and the observations of periodic effects in the previously mentioned nuclides, nearly all of which involve $\beta$-decays \citep{chr83,bun10}. It should be noted, however, that variations in some $\alpha$-decay rates were found in experiments by Shnoll, et al. \citep{shn98a,shn98b}. Moreover, even though periodic effects indicative of a solar influence have been seen in $\beta$-decays, there is no theoretical reason why such effects should have been seen at the same level in all $\beta$-decays. As we have noted elsewhere \citep{fis09,jen10,bun10}, the same properties of $\beta$-decaying nuclei which are responsible for the broad range of observed half-lives (e.g., Q-values, nuclear and atomic wave functions, selection rules), would likely ensure that decaying nuclides would be sensitive in different degrees to an additional small perturbation arising from a common solar influence.

The preceding discussion leads naturally to the analysis of Norman et al. \citep{nor09} which is the focus of the present paper. These authors examined data they had taken for six different isotopes (five $\beta$-decays and one $\alpha$-decay), and found no evidence for the periodic effects reported by other authors (see above) which would signal a correlation between nuclear decay rates and Earth-Sun distance. Even though there is no obvious conflict between the Norman results and the conclusions of Refs. \citep{jen09a,jen09b,fis09,stu10a,stu10b,stu11a,stu11b}, since none of the isotopes studied by Norman et al. were the same as those in which periodic effects have previously been reported, their high-quality data are useful in the present context since they serve to underscore the importance of understanding the phase of nuclear decay data sets exhibiting a periodic time variation. This understanding will lead in turn to more precise limits on the sensitivities of various nuclides to a possible solar influence. (In what follows we use the word ``phase'' to refer to the calendar day on which a maximum in the counting rate is observed in a given data set. 1 January at 00:00 corresponds to phase 0.00, and midnight on 31 December corresponds to phase 1.00).

To understand the question of phase in the context of the Norman analysis, we focus on the decay measurement ratio $^{22}$Na/$^{44}$Ti which Norman et al. exhibit in their Fig. 1 \citep{nor09}. These authors have graciously provided us with their $^{22}$Na/$^{44}$Ti measurements, and we have reproduced their normalized residuals in Fig. 1 after detrending their data to take account of exponential decay. Norman et al. compare these data with two hypotheses: the null hypothesis in which there is no annual variation, and what they call the ``Jenkins hypothesis,'' which is a proposed correlation between their data and $1/R^2$, where $R$ is the Earth-Sun distance. Since $1/R^2$ has a maximum at perihelion, which occurs in the first few days of January (varying from year to year between 2 January and 5 January), the analysis of Norman et al. builds in this phase at the outset even though this is not the observed phase for any of the data sets which exhibit annual variations. For example, the $^{32}$Si/$^{36}$Cl phase in the BNL experiment was 10 February ($\pm$3 days) \citep{alb86,jav10,stu11b}. Hence the actual ``Jenkins hypothesis'' should reflect the observed phase of the existing data, which can be understood as arising from a combination of the annual variation of $1/R^2$ and a possible North-South asymmetry in the solar emission mechanism \citep{stu11b}. The latter effect, which has been ascribed to the tilt of the solar axis of rotation with respect to the ecliptic \citep{stu11b}, produces an additional annual variation with a phase of either 8 March or 8 September. A phase later than perihelion is consistent with the findings of the analyses of data taken at two major neutrino observatories. Analysis of data from Super-Kamiokande-I finds a phase shift from perihelion of approximately 13$\pm$17 days, which puts the peak solar neutrino flux at $\sim$16 January \citep{smy04,hos06}. Data from the Sudbury Neutrino Observatory (SNO) show a more significant shift from perihelion of approximately 40 days \citep{aha05,ran07}, which puts the peak solar neutrino flux at approximately 12 February. We recall that the phase of the BNL $^{32}$Si/$^{36}$Cl data is 10 February ($\pm$3 days) \citep{alb86,jav10,stu11b} and the $^{36}$Cl data taken at the Ohio State Research Reactor exhibits a phase of 17 February ($\pm$3 days)\citep{jen12}.

\begin{figure}[h]
\includegraphics[width=\columnwidth]{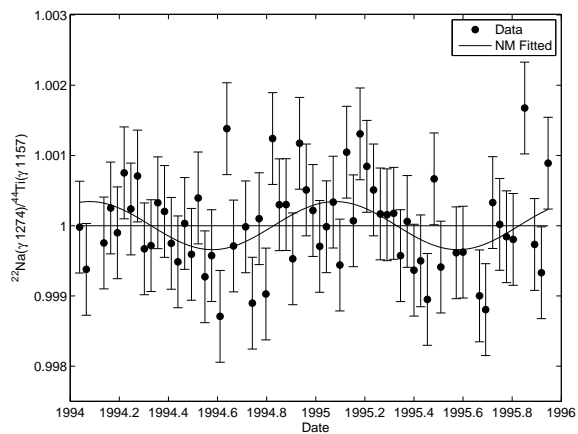}
\caption{Normalized $^{22}$Na/$^{44}$Ti ratio of count rates versus time. The solid black curve is our best sinusoidal fit to the data obtained from the Nelder-Mead (NM) optimization. With the frequency fixed at 1.00/yr, the resulting amplitude is 0.00034, and the phase is 0.08 (29 January). See text for further details. \label{fig:fit}}
\end{figure} 
 
The impact of phase considerations on the analysis of Norman et al. can be understood by a simple example. Consider two sinusoidally varying functions $f_1\left(t\right)=A_1\rm{cos}\left(\omega{}t-\phi_1\right)$ and $f_2\left(t\right)=A_2\rm{cos}\left(\omega{}t-\phi_2\right)$. The correlation coefficient $C\left(f_1,f_2\right)$, defined as the normalized scalar product of $f_1\left(t\right)$ and $f_2\left(t\right)$, is given by

\begin{equation}
C\left(f_1,f_2\right)=\frac{\left\langle{}f_1\vert{}f_2\right\rangle }{\vert{}f_1\vert{}\vert{}f_2\vert }=\rm{cos}\left(\phi_1-\phi_2\right).
\label{eq:eq1}
\end{equation}

\noindent{}It follows from Eq. \ref{eq:eq1} that if we identify $f_1\left(t\right)$ with the data of Norman et al., and $f_2\left(t\right)$ with the ``Jenkins hypothesis'', then the correlation coefficient between  $f_1\left(t\right)$ and $f_2\left(t\right)$ depends on the presumed phase $\phi_2$ of the ``Jenkins hypothesis.'' In the ensuing analysis we allow the phase and amplitude of the Norman data to be free parameters determined by the data themselves. We then demonstrate that in this approach the data of Norman et al. do in fact favor the presence of an annual periodicity in the $^{22}$Na/$^{44}$Ti ratio, which could be attributable to an external influence, possibly the Sun.

%

\section{Power Spectrum Analysis}
\label{psa}
We have analyzed the $^{22}$Na/$^{44}$Ti data provided to us by Norman, et al., using a likelihood procedure \citep{cal05} developed originally for application to the analysis of solar neutrino data. This procedure is similar to the well-known Lomb-Scargle method \citep{lom76,sca82}, but it provides estimates of the amplitudes and phases of oscillations as well as the powers. In contrast to the analysis of Norman et al., where the null hypothesis was compared to an assumed sinusoidal variation whose amplitude and phase were fixed at the outset to be 0.0015 and 3 January, respectively, our analysis allows both of these parameters to be determined by the data themselves.

\begin{figure}[h]
\includegraphics[width=\columnwidth]{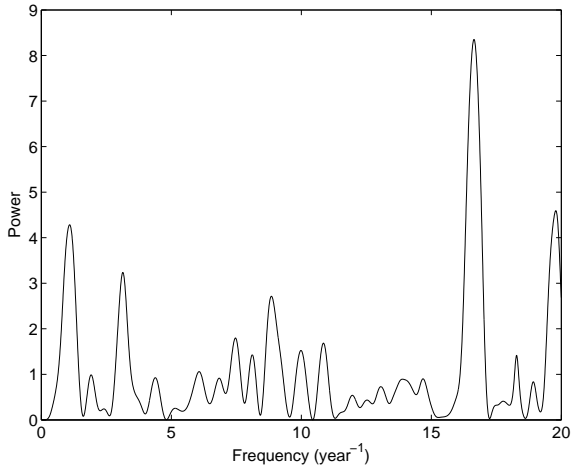}
\caption{Power spectrum of the$^{22}$Na/$^{44}$Ti data of Norman, et al.[31]. \label{fig:ps22na}}
\end{figure}

We formed the power spectrum of the data shown in Figure \ref{fig:fit}, using this likelihood procedure. In a power spectrum analysis, the probability of the null hypothesis (i.e., that the peak is due to normally distributed random noise) is given by $e^{-S}$, where $S$ is the value of the power at a specified frequency. The resulting power spectrum, over the frequency range 0-20 yr$^{-1}$, is shown in Figure \ref{fig:ps22na}, and the strongest peaks in the frequency range 0-20 yr$^{-1}$ are listed in Table 1. The strongest peak (with power $S$=8.35) is found at frequency 16.64 yr$^{-1}$. There is no obvious interpretation of this frequency, but it does hint that there may be unknown influences on the decay process.

\begin{table}
\label{tbl:table1}
\caption{\small{Most prominent frequencies in the range of 0-20 yr$^{-1}$ obtained from the power spectrum analysis of the $^{22}$Na/$^{44}$Ti data of Norman, et al. \citep{nor09}}}
\centering
\begin{tabular}{ l c }
\hline
Frequency (yr$^{-1}$) & Power ($S$)\\
\hline
16.64 & 8.35\\
19.79 & 4.59\\
1.09 & 4.28\\
3.14 & 3.24\\
8.85 & 2.71\\ 
\hline 
\end{tabular} 
\end{table}

\begin{figure}[h]
\includegraphics[width=\columnwidth]{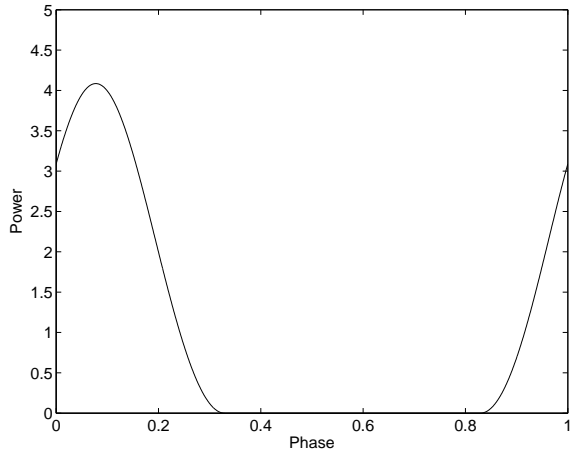}
\caption{Power versus phase for annual oscillation. \label{fig:PvsPhi}}
\end{figure}

The peak at 1.09 yr$^{-1}$ may be identified as an annual oscillation. We find that the amplitude of the normalized data at this frequency is approximately 0.00034, which is smaller by a factor of approximately 4 than the amplitude assumed by Norman et al. (0.0015). This explains why Norman et al. found a closer fit between their data and a flat curve than between the data and a sinusoidal oscillation of amplitude 0.0015. This result also underscores the previous remarks to the effect that there is no theoretical reason to expect that all isotopes should exhibit sinusoidal variations of the same amplitude. We show in Figure \ref{fig:PvsPhi} the power as a function of phase, for a frequency of exactly 1.00 yr$^{-1}$. We find that the peak occurs at an annual phase of 0.08$\pm$0.07, corresponding to the date range 29 January ($\pm$24 d), i.e. to the 1$\sigma$ range 5 January to 22 February. Although this is compatible with the phase assumed by Norman et al. (3 January), it allows a much broader range of phases, which from Eq. \ref{eq:eq1} clearly affects the correlation between their data and the ``Jenkins hypothesis.'' We will return to this point below.
 
\section{Significance Estimate}
\label{sigest}
\begin{figure}[h]
\includegraphics[width=\columnwidth]{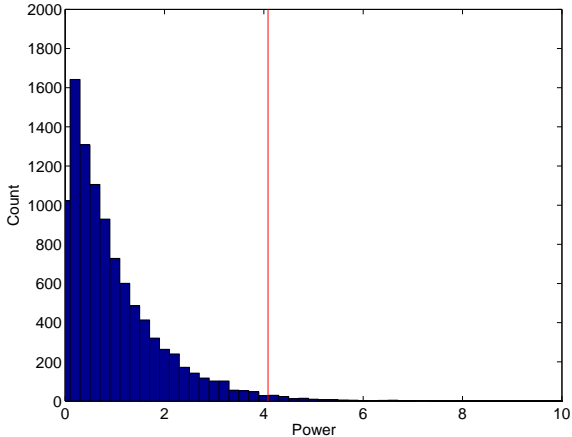}
\caption{Histogram of the power at 1.00 yr$^{-1}$, formed from 10,000 simulations, generated by the shuffle procedure. Only 127 of these simulations have power as large or larger than the power 4.08 of the actual data shown by the vertical line. \label{fig:SigEst}}
\end{figure}

We now estimate the significance of the annual oscillation found by our power-spectrum analysis. The power at exactly 1.00 yr$^{-1}$ is found to be 4.08, and we can estimate the probability that this peak could have arisen by chance by using the standard shuffle test \citep{bah91}. We generated 10,000 simulations of the data by randomly relating the actual measurements and the actual times. The result is shown in histogram form in Figure \ref{fig:SigEst}. We find that, of these 10,000 simulations, only 127 have powers as large or larger than 4.08, the power of the actual data. Hence the annual oscillation revealed by power spectrum analysis appears to be significant at the $\sim$1\% level.
 
\section{Discussion}
\label{disc}
\begin{figure}[h]
\includegraphics[width=\columnwidth]{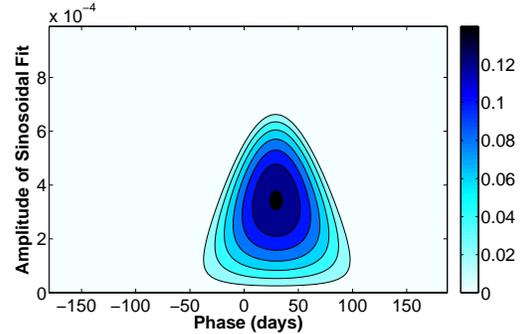}
\caption{Plot of the difference in reduced chi-squared values (null hypothesis  ``Jenkins'') with $f$ = 1.00 yr$^{-1}$, for various amplitudes and phases. Colored regions correspond to sinusoidal parameters where the ``Jenkins hypothesis'' is preferred. The optimal fit, located in the central, darkest region with amplitude $\approx$ 3.4$\times$10$^{-4}$ and phase $\approx$29 d, has a reduced chi-squared difference of $\approx$0.14. 1 January at 00:00 corresponds to phase 0.00, and midnight on 31 December corresponds to phase 1.00.\label{fig:chisq}}
\end{figure}

It follows from the preceding discussion that there is in the end no conflict between the results of our analysis and those obtained previously by Norman et al. The difference between our results and theirs is that we have addressed different questions: Norman et al. asked whether their analysis of a particular dataset (the ratio of $^{22}$Na/$^{44}$Ti) yields the same annual oscillation as might be expected from a phase due to the $1/R^2$ variation of the Earth-Sun distance with an amplitude fixed at the outset by the BNL and PTB data. By contrast, we have asked whether there is evidence for an annual oscillation, for any combinations of amplitude and phase. As shown in our Figure \ref{fig:chisq}, there is a wide range of values for the amplitude and phase for which an annually varying signal provides a better description of the Norman data than does the null hypothesis.
 
We therefore adopt the position that since the mechanism of these oscillations is currently unknown, and since our analyses of several datasets yield evidence of annual oscillations with a range of amplitudes and phases, there is no reason to expect that the amplitude and phase of an annual oscillation in the Norman dataset should be identical to that of the amplitude and phase of either the PTB $^{226}$Ra measurements, or to the ratio of the BNL $^{32}$Si and $^{36}$Cl measurements. This observation can be strengthened by noting that although one can identify an annual oscillation in each of the BNL $^{32}$Si and $^{36}$Cl measurements, the individual phases of each nuclide differ from the phase of their ratio and each other: The phase of the $^{32}$Si measurements is 0.899 (24 Nov) and the phase of the $^{36}$Cl measurements is 0.695 (11 Sep) \citep{stu11b}. This fact indicates that the phase is probably not determined solely by the Earths orbital motion and emphasizes the importance of leaving the phase as a variable to be determined from the data, for each set of measurements. We also note in passing that since the BNL $^{32}$Si and $^{36}$Cl data were acquired at the same time on the same apparatus, the very fact that these data exhibit different phases argues against attributing the observed annual variations in count rates to an environmental influence on the apparatus \citep{jen10}.

Our focus here as been on annual oscillations in some measured nuclear decay data, since these were the only variations discussed by Norman et al. However, these are not the only periodicities which have been observed, and they may not be the most significant. In recent articles \citep{jav10,stu11a,stu11b,stu10b,stu10a,fis11}, we have presented evidence for periodicities in nuclear decays with frequencies in the range 11-13 yr$^{-1}$ and $\sim$2.11 yr$^{-1}$.  The 11-13 yr$^{-1}$ periodicities may be attributable to internal rotation of the Sun. The $\sim$2.11 yr$^{-1}$ periodicities may be an analog of the well-known Rieger oscillation \citep{rie84}, attributable to an r-mode oscillation \citep{sai82} in an inner solar tachocline. It would clearly be of interest to extend this and other analyses of decay data to look for such periodicities which, as noted above, cannot be attributed to known environmental effects. Furthermore, an examination of the literature yields additional evidence for these observed periodicities, annual and otherwise, which have also been found in decay data. A list of the experiments we have found to date is presented in Table \ref{tbl:table2}.

As we show in Table \ref{tbl:table2}, the majority of the isotopes listed have at least one beta-decaying branch, or several beta-decaying daughters. There is one exception, however, and that is the result presented by \citet{shn98a,shn98b}, where an oscillation was observed in $^{239}$Pu, which is strictly alpha-decay through any realizable daughter that could be measured. However, others have made measurements of isotopes which decay by alpha with no realizable beta-emitting daughters and have seen no such effects, e.g. the results of \citet{nor09} and \citet{par10a}. The mechanisms of beta- and alpha-decays are obviously very different. While we still have not determined a mechanism by which solar neutrinos would affect the weak interaction associated with beta-decays, the development of a model where the neutrinos could affect both alpha- and beta-decays becomes even more difficult. We note, however, that there are thousands of isotopes of varying decay modes which have not been examined. Many of these measurements will be complicated due to very short half-lives, or other properties, which may make many measurements impractical with present technologies. This includes the beta- and gamma-decays of all known elementary  particles, including the neutron, and many nuclides. However, there remain many avenues to explore in this work.

\acknowledgments

We are indebted to E.B. Norman and his collaborators for generously sharing with us the data analyzed in this article. The work of P.A.S. was supported in part by the NSF through Grant AST-06072572, and that of E.F. was supported in part by U.S. DOE Contract No. DE-AC02-76ER07128. The views expressed in this paper are those of the authors and do not reflect the official policy or position of the U.S. Air Force, U.S. Department of Defense, or the U.S. Government.  The material in this paper is Unclassified and approved for public release. Distribution is unlimited, reference Air Force Flight Test Center Public Affairs.

\bibliographystyle{spr-mp-nameyear-cnd}

\clearpage

\begin{deluxetable}{ccccll}
\tabletypesize{\footnotesize}
\tablecaption{Some experiments where time-dependent decay rates have been observed.\label{tbl:table2}}
\tablewidth{0pt}
\tablehead{
 & \colhead{Decay} & \colhead{Detector} & \colhead{Radiation} &  \\
\colhead{Isotope} & \colhead{Type} & \colhead{Type} & \colhead{Measured} & \colhead{Reference}
}
\startdata
$^{3}$H & $\beta^{-}$ & Photodiodes & $\beta^{-}$ & \cite{fal01}  \\
$^{3}$H & $\beta^{-}$ & Liq. Scint. & $\beta^{-}$ & \cite{shn98a,shn98b}  \\
$^{3}$H & $\beta^{-}$ & Liq. Scint. & $\beta^{-}$ & \cite{vep12}  \\
$^{3}$H & $\beta^{-}$ & Sol. St. (Si) &$\beta^{-}$ & \cite{lob99} \\
$^{22}$Na/$^{44}$Ti\tablenotemark{a} & $\beta^{+},\kappa$ & Solid State (Ge) & $\gamma$ & \cite{nor09} and this article \\
$^{36}$Cl & $\beta^{-}$ & Proportional &$\beta^{-}$ & \cite{jen09b,stu10a,stu11a} \\
$^{36}$Cl & $\beta^{-}$ & Geiger-M\"{u}ller  &$\beta^{-}$ & \cite{jen12} \\
$^{54}$Mn & $\kappa$ &  Scint.  & $\gamma$ & \cite{jen09a} \\
$^{54}$Mn & $\kappa$ &  Scint.  & $\gamma$& \citep{jen11}\\
$^{56}$Mn & $\beta^{-}$ &  Scint. &$\gamma$ & \cite{ell90} \\
$^{60}$Co & $\beta^{-}$ & Geiger-M\"{u}ller  &$\beta^{-}$,$\gamma$ & \cite{par10a,par10b} \\
$^{60}$Co & $\beta^{-}$ & Scint. &$\gamma$ & \cite{bau07} \\
$^{85}$Kr & $\beta^{-}$ & Ion Chamber &$\gamma$  & \cite{sch10} \\
$^{90}$Sr/$^{90}$Y & $\beta^{-}$ & Geiger-M\"{u}ller  &$\beta^{-}$ & \cite{par10a,par10b,stu12a} \\
$^{108m}$Ag & $\kappa$ & Ion Chamber &$\gamma$  & \cite{sch10} \\
$^{133}$Ba & $\beta^{-}$ & Ion Chamber & $\gamma$ & Jenkins, et al., arXiv \\
$^{137}$Cs & $\beta^{-}$ & Scint. & $\gamma$ & \cite{bau07}\\
$^{152}$Eu & $\beta^{-},\kappa$  & Sol. St. (Ge) & $\gamma$\tablenotemark{b} & \cite{sie98} \\
$^{152}$Eu & $\beta^{-},\kappa$ & Ion Chamber &$\gamma$ & \cite{sch10} \\
$^{154}$Eu & $\beta^{-},\kappa$ & Ion Chamber &$\gamma$ & \cite{sch10} \\
$^{222}$Rn\tablenotemark{c} & $\alpha,\beta^{-}$  & Scint. &$\gamma$ & \cite{ste11,stu12b} \\
$^{226}$Ra\tablenotemark{c} & $\alpha,\beta^{-}$ & Ion Chamber &$\gamma$ & \cite{jen09b,stu10b,stu11a} \\
$^{239}$Pu & $\beta^{-}$ & Sol. St.  &$\alpha$ & \cite{shn98a,shn98b} \\
\enddata
\tablenotetext{a}{Only the count rate ratio data were available.}
\tablenotetext{b}{Only the $\kappa$ photon was measured.}
\tablenotetext{c}{Decay chain includes several primarily $\beta$-decaying daughters which also emit photons.}
\end{deluxetable}

\end{document}